The manuscript on the following 2 pages is designated for the

7th Triennial Special Issue of the IEEE Transactions on Plasma Science

"Images in Plasma Science"

scheduled for publication in August 2014

The manuscript length is limited to 2 pages only. It was submitted for peer review on 2013-11-21. Receipt was acknowledged on 2013-11-21 and the manuscript number **TPS7217** was assigned.

# Drifting Ionization Zone in Sputtering Magnetron Discharges at Very Low Currents

André Anders, Pavel Ni, and Joakim Andersson

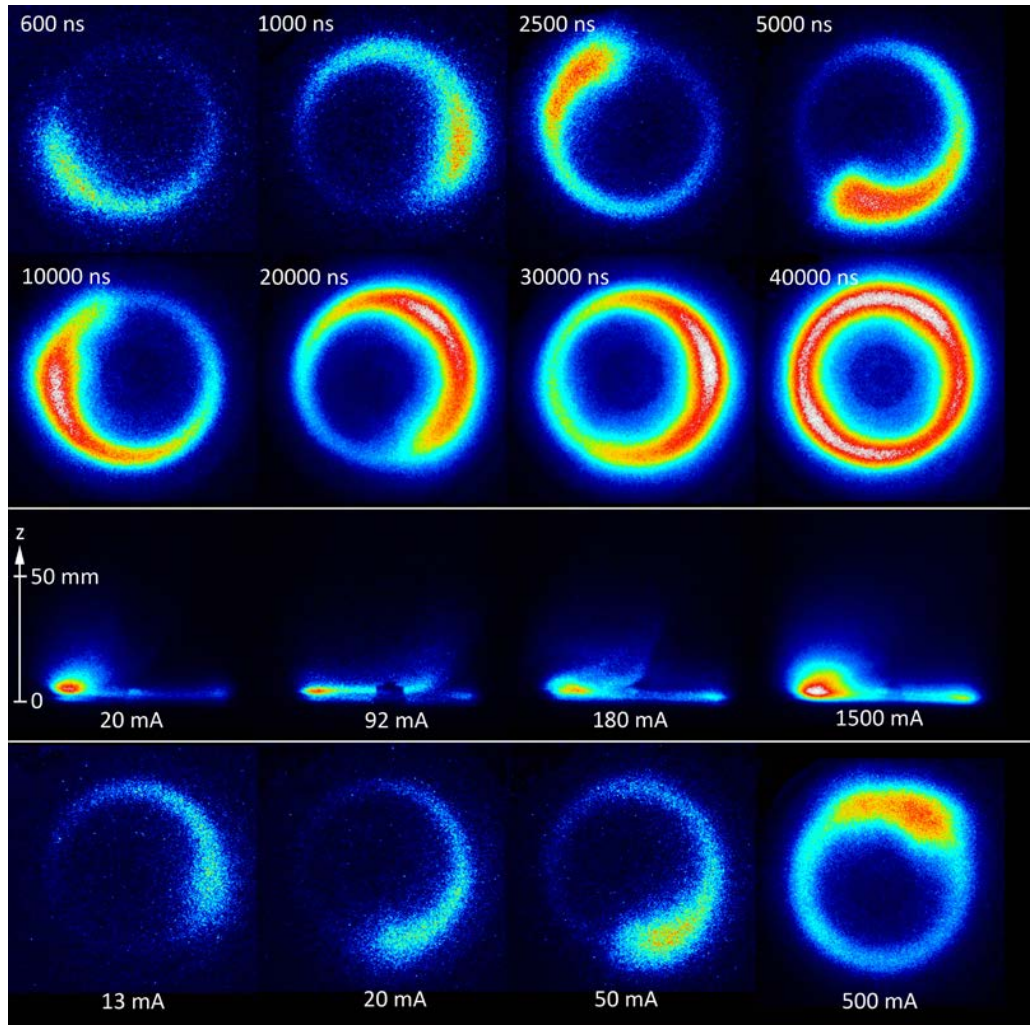

Fig. 1. False-color fast camera images of aluminum sputtered in 0.5 Pa of argon. The top 8 images show the blurring of a drifting ionization zone when the image exposure time exceeds 5 µs (the discharge current was 30 mA). The center 4 side-view images were taken with an exposure time of 500 ns for different discharge currents; $z$ is the distance from the target surface. The lower 4 images, also taken with 500 ns exposure time, illustrate the existence of one zone at different currents. Note that the camera was always in the linear regime; the image intensifier's gain had to be increased to make plasma visible at short exposures, and decreased to avoid saturation at long exposures, and hence the apparent brightness should be judged with caution.

*Abstract* - **Discharges with crossed electric and magnetic fields are known to develop instabilities that are crucial in the transport of charged particles. Sputtering magnetrons are no exception. While most recent studies focused on traveling ionization zones in high power impulse magnetron sputtering, we show here fast camera images of magnetron discharges at very low current. A single drifting ionization zone is always present, even down to the threshold current of about 10 mA.**

Manuscript received …. November 2013.
A. Anders and Pavel Ni are with the Lawrence Berkeley National Laboratory, 1 Cyclotron Road, MS 53, Berkeley, California 94720, USA;
J. Andersson is with the Centre for Quantum Technologies, National University of Singapore, 3 Science Drive 2, 117543 Singapore.
Work at LBNL was supported by U.S. Department of Energy under Contract No. DE-AC02-05CH11231. J.A. acknowledges support from the National Research Foundation and the Ministry of Education, Singapore.
Publisher Identifier S XXXX-XXXXXXX-X

A large class of discharges makes use of electron trapping in crossed electric and magnetic fields, for example Hall thrusters [1, 2], magnetized microdischarges [3], and sputtering magnetrons [4]. It has been recognized that ionization and particle transport are governed by a host of waves and instabilities. Recent publications on instabilities in magnetrons focused on high power impulse magnetron sputtering (HiPIMS); several groups showed pronounced plasma non-uniformities drifting along the "racetrack." These regions of excitation and ionization have been called ionization zones [5, 6], bunches [7], spokes [8], or emission structures [9].

In this contribution, we use fast camera imaging to study plasma non-uniformities in direct current (dc) magnetron discharges in end-on and side-on view. While it is well established that HiPIMS discharges show strong non-uniformities, the issue is less clear for dc operation. It is more challenging to take images with high temporal, spatial, and spectral resolution given the much weaker light emission.

Different target materials have been used but we limit this report exemplifying the results using an Al target 76 mm (3 inch) in diameter, and 6.2 mm (1/4 inch) in thickness. Sputtering was done in pure argon with a flow rate of 60 sccm resulting in a pressure of 0.5 Pa. The choice of aluminum allows us to do a direct comparison with images obtained in a recent study of HiPIMS discharges [10]. In fact, the target and magnetron were the same, and the power supply (SIPP by Melec) was simply switched to dc operation. The current was measured using a Fluke 189 digital multimeter, and the voltage was checked at the current feedthrough using a $100\times$ voltage divider (Tektronix P5100).

The images were taken with a Princeton Instruments PI-MAX 1024 camera equipped with a microchannel plate image intensifier. The sensitivity range of the photodetector was from about 290 nm to about 810 nm. The exposure time could be reduced to 1 ns but most images were taken with a much longer exposure to (a) improve the signal-to-noise ratio, and (b) determine the approximate drift velocity of one or more luminous regions, should they be present, by using various exposure lengths. The intensities are presented in false color using the color scale "royal" of the image processing software ImageJ [11].

Figure 1 shows selected images obtained with different exposure times, discharge currents, in end-on and side-on view, as indicated by the labels and in the figure caption. We see that no significant blurring occurs until the exposure time exceeds 5 µs. This is in contrast to HiPIMS, where the exposure time had to be shorter than 1 µs to avoid blurring [6]. In dc sputtering at low current (1.5 A or less), we always see a single zone, not several as it is typical in HiPIMS. The time for one rotation is 60 µs or more (at 30 mA current), giving a zone drift velocity of 2000 m/s or less, which is about 20% or less than the zone drift velocity in HiPIMS [6]. Unfortunately, our single-image technique does now allow us to make a statement about the direction of the zone drift relative to the $\mathbf{E}\times\mathbf{B}$ drift of electrons.

Staying with a short exposure of 500 ns, the drifting ionization zone is imaged at very low currents. The driving voltage was stepwise reduced from 295 V (500 mA) until the discharge extinguished at about 245 V (13 mA). Even at the smallest current, one ionization zone could be observed: there is no threshold for its formation.